\documentclass[preprint]{aastex}
\usepackage{natbib}
\bibpunct[;]{(}{)}{;}{a}{}{,}

\journalid{337}{15 January 1989}
\articleid{11}{14}

\begin{document}

\title{Evidence for a Young Stellar Population in NGC~5018}

\author{Andrew J. Leonardi\altaffilmark{1} }
\affil{CB \#3255, Department of Physics \& Astronomy, University of North
    Carolina, Chapel Hill, NC 27599-3255}

\and

\author{Guy Worthey}
\affil{Department of Physics \& Astronomy, St. Ambrose University, 
518 W. Locust St., Davenport, IA 52803-2829}


\altaffiltext{1}{Visiting Astronomer, Cerro Tololo Inter-American Observatory. 
CTIO is operated by AURA, Inc.\ under contract to the National Science
Foundation.} 


\begin{abstract}
Two absorption line indices, Ca II and H$\delta$/$\lambda$4045, measured
from high-resolution spectra are used with evolutionary synthesis
models to verify the presence of a young stellar population
in NGC~5018.  The derived age of this population is $\sim$ 2.8 Gyr with a 
metallicity roughly solar and it completely
dominates the integrated light of the galaxy near 4000 \AA. 
\end{abstract}


%
%

\section{Introduction}

Morphological peculiarities in the optical images of galaxies are now almost
invariably taken to be signs of a past tidal interaction or merger.  Though
theoretical models have had success reproducing tidal tails, shells, and
other structures \citep[e.g.,][]{tt72,q84,hq88,mbr93},
in the absence of explicit details of the interaction such as the Hubble
types of the progenitors, relative progenitor sizes, and impact parameter, a 
unique solution is difficult to come by.

Dynamical friction during a collision almost certainly leads to a merger of the
stellar systems \citep{s83} and simulations have shown that any
accompanying gas will rapidly dissipate to the center during minor mergers involving
either ellipticals \citep{wh93} or disk galaxies
\citep{mh94a}.  If the conditions are right, it is
reasonable to expect that this gas inflow will result in star formation.  The
duration, intensity, and even the starting time of the star formation, however,
depend on the morphology of the progenitor and the details of the
interaction \citep{mh94b}.  As much post-merger 
information as possible is needed to help constrain the merger possibilities, 
including analysis of the resulting young stellar population (YSP).

Unfortunately, unless the merger remnant is presently forming stars at a 
reasonably vigorous rate (hence producing readily apparent emission lines) or has
experienced a very recent and relatively strong episode of star formation placing
it in the starburst regime, a YSP can be difficult to detect.  Broadband colors
and other YSP indicators return to pre-star formation levels very quickly
\citep{bas90,cs94} and
many indicators that imply the presence of a YSP can also be explained by
intrinsic metallicity differences in the final population of the merger
remnants \citep[hereafter BBB]{bbb93}.  Ambiguity
about whether a YSP even exists or not complicates the details of the merger
considerably.

In this paper, we use spectral indices in conjunction with evolutionary synthesis models to 
detect and determine the age and metallicity of a YSP in the particular case of
a possible merger remnant,
NGC~5018.  It represents an update to the age-dating technique introduced in
\citet[hereafter LR]{lr96}.  NGC~5018 is appropriate to the
present discussion because there is ongoing uncertainty concerning the presence of
a YSP.  Certain observations imply the existence of a YSP while others seem to be
inconsistent with the presence of a YSP.  NGC~5018 and the roots of the controversy 
are described
in \S2.  In \S3, a review of the age-dating technique and its 
refinements are given.  The results of the technique applied to NGC~5018 are
presented in \S4 and \S5 contains the conclusions.

\section{NGC~5018}

NGC~5018 is a member of the \citet{mc83} catalog of
shell elliptical galaxies and is considered a probable merger remnant
\citep{fpcmv86}.  As noted by \citet{ssfbdg90}
and BBB, NGC~5018 has an abnormally weak $\mathrm{Mg_{2}}$ index
for its luminosity: Its measured Mg$_2$ is 0.209 \citep{twfbg98}
even though the mean Mg$_2$-$\sigma$ relation suggests Mg$_2 = 0.301$
for an elliptical galaxy with NGC 5018's measured velocity dispersion
of $\sigma = 223$ km s$^{-1}$ \citep{bbf93}, about 6 standard
deviations away from the mean. 
Although deviations from the line-strength-luminosity
relation correlate well with the amount of morphological disturbance in
elliptical galaxies \citep{ssfbdg90}, NGC~5018 has
abnormally weak line strengths even when this correlation is accounted for.
For a class of objects, these authors ruled out metallicity
variations in the galaxies as the cause for the correlation due to the 
physical implausibility
of stronger mergers leading to more metal-poor stellar populations in the
remnants.  \citeauthor{ssfbdg90} concluded that mergers produce a YSP which is
observed in the decreased line strengths.  On a galaxy-by-galaxy basis,
however, an intrinsic metallicity variation cannot be ruled out by
a low $\mathrm{Mg_{2}}$ index alone. In NGC~5018, a low $\mathrm{Mg_{2}}$ 
index coupled with the lack of an upturn
in its UV spectral energy distribution (SED), led BBB to conclude that
NGC~5018 consisted of a metal-poor old stellar population, in stark
contrast to other ellipticals of the same luminosity.
BBB were unable to match both the UV SED observations and the $\mathrm{Mg_{2}}$
index with composite populations created by mixing spectral templates of
metal-rich elliptical galaxies and a contaminating YSP template.  Only
templates containing metal-poor populations approached both observations.

Indirect observational evidence that a YSP does in fact exist in NGC~5018 is
extensive.  The shells present in its optical image are photometrically
bluer than the surrounding parts of the galaxy \citep{fpcmv86}
suggesting a younger age for the shells.
The detection of an HI gas bridge connecting NGC~5018 with the nearby spiral
NGC~5022 \citep{kgvjk88} is evidence of an ongoing interaction
while a possible past interaction is implied by a stellar
bridge connecting the two and the embedded dust lane in NGC~5018
\citep{mh97}.
Possible young globular cluster candidates, formed during a past interaction
and perhaps only several hundred Myr old, have been observed 
\citep{hk96}.
Furthermore, \citet{ghjn94} measured extended H$\alpha$+[N II]
emission in the central region coinciding with the embedded dust lane which
they associated with star forming regions.
Also, IR emission has been detected in the same area 
\citep{jkkg87}. 
\citet{tb87} showed in an IR two-color diagram that
NGC~5018 lies in a region quite different
from that occupied by infrared ``cirrus'', which is emission from diffuse
dust in the interstellar medium of a galaxy. Instead, it is closer to
the region where the IR emission from warmer dust associated with HII 
regions dominates \citep{h86,blw88}
, suggesting a YSP source for the IR emission.

While the observations are compelling, they are not conclusive.  The
direct detection of the YSP is needed to resolve the issue.  BBB chose
to observe in the far-UV for exactly that reason, since in principle,
this region of the spectrum is dominated by young stars
\citep{o88} and the low UV flux level in NGC~5018 led
them to the metal-poor scenario.  They discounted dust obscuration as
the cause because the best available photometry at that time
\citep{fpcmv86} showed that the dust lane in NGC~5018
does not extend into the region where their IUE spectrum was taken.
Subsequent observations, however \citep{cd94,ghjn94}
, indicate that not only is dust
present throughout the central region but also is patchy in nature,
making the reddening effects difficult to ascertain.
\citet{cd94} showed that reasonable
expectations of dust obscuration and a YSP can explain both the low
$\mathrm{Mg_{2}}$ index and the UV flux depletion in the central
region of NGC~5018.  Both sets of authors concluded that a YSP was a
more probable explanation for the observations.  

The conflict is
illustrated in Figure~\ref{fig:lick} where we have plotted data from
the Lick group \citep{twfbg98} for NGC~5018 and other
systems.  The top panel shows a $\lambda 2750-V$ color plotted against
the $\mathrm{C_{2}}$4668 Lick index. The $\mathrm{C_{2}}$4668 index is
more metal-sensitive than Mg$_2$ and is less subject, but not immune,
to abundance ratio effects.  BBB remarked that NGC~5018's
$\mathrm{Mg_{2}}$ index and UV spectrum resembles M32's, even though
M32 is a much less luminous galaxy, which led them to surmise a system
with an abundance like that of a dwarf galaxy rather than a giant
elliptical. The UV data here supports this view.  As can be seen in
the top panel of Figure~\ref{fig:lick}, NGC~5018 has approximately the
same UV color but a slightly weaker $\mathrm{C_{2}}$$\lambda$4668 index than M32
indicating that NGC~5018 is about twice as old and more metal-poor
than M32.  The bottom panel of Figure~\ref{fig:lick} plots H$\beta$
(uncorrected for any emission fill-in) against $\mathrm{C_{2}}$4668
and shows NGC~5018 nearly on the same model grid line as M32 implying
similar age, but still $\sim$ 0.15 dex more metal-poor than M32.
Emission corrections for H$\beta$ would push NGC~5018 vertically
upward to younger ages and higher metallicity, worsening the
discrepancy in age between the two panels. On the other hand, a
correction for UV extinction in the upper panel would make the two
panels agree better.

To help resolve the still uncertain nature of NGC~5018, we utilize 
spectroscopic observations, along with an updated age-dating method to
unambiguously show that a YSP is present in NGC~5018.
The modeling technique is discussed in \S3.  The 
long-slit spectra of NGC~5018 were acquired at the KPNO 4m telescope in June~1995 by
Lewis Jones and kindly provided to us.  Four 30-minute exposures were acquired
with the R-C spectrograph with grating KPC-22B at second order and the T2KB
2048x2048 CCD.  The slit width was 2 arcsec.  A 14.5 pixel aperture was extracted
from the raw spectrum and with a CCD spatial scale of $\sim$ 0.69 arcsec/pixel,
the aperture size on NGC~5018 is 2''x10''.  The dispersion of the spectra is 
0.7 \AA/pxl and the resolution
is FWHM $\sim$ 1.8 \AA.  Data reductions were done in IRAF.  For details see
\citet{j99}. A representative spectrum, emphasizing the wavelength
region of interest, is shown in Figure~\ref{fig:spec}.  It has been normalized
to unity at 4040 \AA.

\section{The Age-Dating Technique}

The age-dating technique as described in LR uses two spectral indices developed
in \citet{r84, r85}.  Each index is defined by taking the ratio
of counts in the bottoms of two neighboring absorption lines without reference
to the continuum levels. The specific absorption lines used are identified on the
NGC~5018 spectrum in Figure~\ref{fig:spec}. The first index, H$\delta$/$\lambda$4045, 
measures
the integrated spectral type of a galactic stellar population and is produced
from the ratio of the central intensity in H$\delta$ relative to the central 
intensity in the neighboring
Fe I $\lambda$4045 line.  Note that the way the index is defined, 
H$\delta$/$\lambda$4045 \emph{decreases} as H$\delta$ gets stronger.  The second 
index, Ca II, formed from the ratio of
the central intensity of Ca II H+H$\epsilon$ relative to Ca II K, is constant in
stars with a spectral type later than F2 but then decreases dramatically for 
earlier type stars, reaching a minimum at spectral type A0.  It provides an
unambiguous signature for stars hotter than F2 in the integrated light of a 
stellar population.  

The indices are computed for single-age theoretical populations from evolutionary
synthesis models.  When plotted together in the two-dimensional index space, the
indices resolve the well-known degeneracy between a YSP's age and light 
contribution to a composite stellar population \citep[e.g.,][LR]{cs87,bas90,cs94}.
A valuable property of these indices particularly applicable to 
NGC~5018 is their virtual insensitivity to reddening.  Since only neighboring
spectral features are used, reddening does not affect their values and the
embedded dust in the center of NGC~5018 will not obscure the evidence of a
YSP.

In LR, the evolutionary synthesis models of \citet{bc93}
were used with the spectral library updated to include the higher resolution
stellar library of \citet{jhc84}.  The models
were restricted to solar abundance populations only, thus metallicity effects on the 
indices could only be explored crudely.  To remedy this, the technique now employs
the evolutionary synthesis models of \citet{w94}.

The Worthey models work as follows:  For a given age and metallicity, a theoretical
isochrone \citep{bbcfn94} is consulted, with each point on
the isochrone representing a parcel of stars of known luminosity, temperature, 
and gravity.  The spectral indices for that isochrone point are interpolated
from empirical fitting functions of the indices from a high-resolution spectral
library \citep[see also \citealt{lea96}]{j99} that has been 
smoothed to the resolution of the
NGC~5018 spectra.  The indices are weighted by luminosity and
number and added up along the isochrone to get the spectral index values for the
entire integrated population.  The population is formed from an instantaneous
burst of star formation.  A finite burst, however, is more realistic and will become
a future feature of the models.  These models were originally designed to disentangle
age and metallicity effects in the integrated light of old stellar populations.  To
extend the age coverage to include very young ages ($<$ 1 Gyr) and also extend spectral
coverage to the blue, the empirical library was augmented with 2103 
theoretical stellar spectra computed with the \citet{k95} SYNTHE 
program.  The indices were computed for each synthetic spectrum and the values were
used as a lookup table with specific isochrone points being found by interpolating
between the synthetic grid points \citep[for more details see][]{l00}.

Figure~\ref{fig:burst} shows the Ca II index plotted against the 
H$\delta$/$\lambda$4045 index
for two stellar populations with [Fe/H] = -0.7 and [Fe/H] =
0.0 respectively. 
In the figure, the solid squares represent the index values for an instantaneous
burst of star formation that has evolved to the labeled age in Gyr, so each curve
follows the evolution of the indices for a stellar population of
the given metallicity.  Both indices initially decrease for young systems as they 
age, reaching a minimum at about 0.25--0.5 Gyr as the O and B stars die
out and A stars begin to dominate the integrated light, thus generating strong
Balmer lines.  Subsequently, as the Balmer lines weaken, the indices increase
again. Also plotted in Figure~\ref{fig:burst} are 
the index values observed
for a select Galactic globular cluster, 47 Tuc ([Fe/H] = -0.7).  
The long-slit observation was  acquired at the CTIO 1.5m telescope in November 
1995.  A 10 minute exposure of 47~Tuc was acquired using the Cassegrain spectrograph
with the Loral 1200x800 CCD and the B\&L grating \#58 at second order.  During the 
exposure, the slit was trailed across the core diameter of the cluster to obtain
a true integrated light
measurement.  The dispersion of the spectrum is 1.12 \AA/pixel and the resolution is
FWHM $\sim$ 2.6 \AA.  Data reductions were done in IRAF.  For details see 
\citet{l00}.  Although the Ca II index loses much of its
age discriminating power at older ages, we still determine a reasonable 
globular cluster age of approximately~15 Gyr for 47~Tuc at the appropriate
metallicity. Simultaneous age and metallicity discrimination with high S/N spectra is 
most effective between
the ages of 0.25 Gyr and 4 Gyr. Since the Ca II and H$\delta$/$\lambda$4045 
indices are insensitive to metallicity for ages less than $\sim$ 0.25 Gyr, the ability to 
uniquely determine a metallicity is lost whereas for ages greater than $\sim$ 4.0 Gyr, the
Ca II index approaches the constant value for late type stars and its evolution essentially
halts.

Unfortunately, the isochrones used here only model the horizontal branch as a red
clump.  
To illustrate where such a population with a blue horizontal branch would fall on the
Ca II--H$\delta$/$\lambda$4045 diagram, also included
on Figure~\ref{fig:burst} are the index values for a 20 minute
exposure of the Galactic globular cluster M15 taken during the same observing 
run as 47 Tuc and for a very metal-poor ([Fe/H] = -1.7) 15.1 Gyr model with
a red clump (note that both the models in 
Figure~\ref{fig:burst} and the globular cluster spectra 
have been smoothed out to the resolution and intrinsic velocity dispersion of 
the NGC~5018 spectra described 
above).  While~47 Tuc has a predominantly red horizontal branch, M15 has a blue
horizontal branch \citep{l89}.  The agreement between the [Fe/H] = -1.7
model and M15
is poor.  Blue horizontal branch stars are luminous enough to contribute significantly
the Ca II 
index and drive it below what the models predict for this age and metallicity.  

\section{Results}

\subsection{Composite Populations}

The question we must answer for NGC~5018 is whether a YSP is contaminating the
light of an old, underlying population or the light is originating from a solely
old, metal-poor population.  To create a composite population, we assume that the
old population has an age of 15 Gyr but the metallicity can vary.  We then 
interpolate in the index space between the old population point and a YSP point in 
constant increments to represent different levels of contamination by the YSP.
The flux
contributions for the two populations have been normalized at 4040 \AA.
The computed indices for the theoretical composite populations are illustrated
in Figures~\ref{fig:yspm17}--\ref{fig:yspp00}.  In each figure, the full
evolution of the YSP is plotted in a similar manner to Figure~\ref{fig:burst},
one YSP metallicity per figure.  A solid, colored circle represents the index
values for the~15 Gyr population denoting the old underlying stellar
system, each color a different old population metallicity which may or may
not be the same as the YSP's metallicity.  The lines connecting the old
population points with select ages along the YSP evolution curve represent
composite stellar populations.  The crosses along each line are interpolations
between these two populations in 25\% increments
of the contribution of the YSP to the
integrated light near 4000 \AA. The YSP points used for interpolation were chosen
so that the resulting composite population would come as closely as possible to
NGC~5018.

The mean indices for the four spectra of NGC~5018 are plotted as an open triangle 
with error bars in Figures~\ref{fig:yspm17}--\ref{fig:yspp00}. Error bars for the 
NGC~5018 data points were calculated by computing the rms scatter among the four
observations. The model trajectories were computed after both the empirical and the 
synthetic spectral libraries
had been smoothed with gaussians to match the resolution
and intrinsic velocity dispersion of the NGC~5018 spectra which allows comparison between
the theoretical integrated indices and those of NGC~5018.
The observed indices for~47 Tuc and a spectrum of M32, obtained during the same observing
run as the globular cluster spectra, are also plotted in 
Figures~\ref{fig:yspm17}--\ref{fig:yspp00} for comparison purposes.

\subsection{Index Plots}

Figure \ref{fig:yspm17} shows the index values for a very metal-poor
YSP ([Fe/H] = -1.7) mixed with two possible old, underlying
populations.  Figure~\ref{fig:yspm07} does the same for a moderately
metal-poor YSP ([Fe/H] = -0.7).  If we also allow the old population
to be metal-poor (red symbols), we have an extreme version of the case
put forth by BBB of a strictly metal-poor population.  If NGC~5018 were to
have such a
population and yet still contain the large amount of structure
attributed to the galaxy by \citet{ssfbdg90}, BBB
postulated that NGC~5018 would either have to be the result of the merging of many
metal-poor components or a large metal-poor elliptical
that experienced a recent merger.  The figures show quite definitively
that an old stellar population with a globular cluster-like metallicity of -1.7 combined
with either the very metal-poor YSP or the moderately metal-poor YSP is 
disallowed. Even by choosing
an age of 13.2 Gyr for the ``young" population to approach NGC~5018
as close as possible, the set of allowed indices
defined by the possible composite populations are not near the location of 
NGC~5018 in the figures.  No combination
of metal-poor young and old populations nor a single coeval metal-poor
population can reproduce the observed indices.

If a metal-poor YSP is mixed with a metal-rich old population, we have the
situation depicted with the blue symbols in Figures~\ref{fig:yspm17} 
and~\ref{fig:yspm07}.  In this scenario, NGC~5018 could
have evolved as a normal elliptical galaxy but then interacted with a 
young, metal-poor disk galaxy.  Dust obscuration as suggested by
\citet{cd94} would still be needed to explain the
lack of an upturn in the UV SED. In any case, this population mixture
is disallowed as well, in agreement with BBB.  If a large percentage
of the light is originating in the old population, the composite has a
similar Ca II index value as NGC~5018 but its H$\delta$/$\lambda$4045
index is too weak compared to the dominant population in NGC~5018.

It is only when we allow the YSP to be metal-rich, i.e., solar or greater,
that the model indices match the observed values for NGC~5018.  
Figure~\ref{fig:yspp00full} shows the solar YSP curve from Figure~\ref{fig:burst}
again along with four old population points, two metal-poor (green and red
symbols) and two metal-rich (blue and magenta symbols).  For clarity, the
interpolation curves between the old population points and the YSP curve have
been omitted from Figure~\ref{fig:yspp00full} and are instead shown on
Figure~\ref{fig:yspp00} which is identical to Figure~\ref{fig:yspp00full} but
on an expanded scale.  We derive an age of $\sim$ 2.8 Gyr for the YSP in
NGC~5018.  Also it is interesting to note in Figure~\ref{fig:yspp00} that the
light at 4000 \AA\  is completely dominated by the YSP with virtually 100\% of the
light originating from the YSP regardless of the assumed old population metallicity.
In fact, for an old population with [Fe/H] = -1.7, \emph{any} contribution
to the light by the old population will drive the model indices away from
NGC~5018. Even in the most likely scenario, a metal-rich old population with a
metal-rich YSP, the old population does not contribute to the integrated light.
In this picture, NGC~5018 can evolve as a normal elliptical galaxy and then 
interact with a  metal-rich companion requiring no unusual events to create the stellar
population of NGC~5018.
 \emph{Thus if an underlying old metal-poor population is present
in NGC~5018, it cannot be contributing significantly to the integrated blue
light.}  

We conclude on the basis of these results that there is a young stellar
population in the central regions of NGC~5018.  We infer that the age of
this YSP is on the order of 2.8 Gyr, the metallicity is near
solar, and it is providing virtually all of the light at 4000 \AA.  These
determinations are reddening independent due to the nature of the spectral
indices used.

\section{Discussion}

The two observations of BBB which led them to conclude that a YSP is
not present in NGC~5018 are a low $\mathrm{Mg_{2}}$ index coupled with
the lack of an upturn in the UV SED both comparable to that found in M32. As seen
in Figures~\ref{fig:yspm17}--\ref{fig:yspp00}, however, NGC~5018 lies in a 
different region of the H$\delta$/$\lambda$4045--Ca II diagram than M32 hence
it cannot be just a high luminosity version of M32.  With the patchiness of 
the dust in the central regions of NGC~5018, there may be as much as two
magnitudes of extinction in the far-UV part of the spectrum which
could explain the lack of an upturn 
\citep{cd94}.  The results
derived here, however, indicate that large amounts of dust obscuration
need not be invoked, although a modest amount of extinction would
bring Figure \ref{fig:lick} into better harmony with the spectral
index results.  The YSP age of~2.8 Gyr derived here is reddening
insensitive and thus robust regardless of the structure of the dust
distribution.  Also, as Figure~\ref{fig:yspp00} shows, the YSP is
completely dominating the visible part of the spectrum of NGC~5018.  A
YSP of this age does not have an upturn in the UV part of the spectrum.
Hence the small far-UV upturn in NGC~5018 may simply reflect the
characteristics of the 2.8-Gyr-old stellar population
dominating the light.

The models predict that a YSP of this age is about 6 times brighter (at
4000\AA ) than an ancient population of the same metallicity and IMF. Hence if
the relative contributions from the old and young populations were only half
and half, then the galaxy must be 6 parts old to 1 part
young in the central region.  This would qualify as a gas-rich
merger event with a rapid gas inflow to the center \citep{wh93}
 of NGC 5018.  If the light contribution is
weighted even more heavily toward the young population as we suggest, this
implies a more violent event, perhaps a merger of two spiral 
galaxies or some other event involving large amounts of gas consumed in
star formation, with an extremely high inflow to the nucleus. Note that,
if nebular emission is filling in the Balmer lines, the derived age
decreases, the YSP is brighter, and so the size of the merging event
decreases.

A solar-[Fe/H] population of age 2.8 Gyr has an Mg$_2 \approx 0.20 $ mag,
as is observed in NGC~5018. Aging this population to 15 Gyr raises the
Mg$_2$ to 0.27 mag, still about 2 standard deviations from the mean
relation for elliptical galaxies, though a dust-hidden YSP could be diluting
the Mg$_2$ index \citep{cd94}. We note, however, 
that compared to
M32, NGC~5018's C$_2$4668 index is weaker but its Mg$_2$ index is
stronger, implying that NGC~5018 may also participate in the general
trend for large ellipticals to have enhanced Mg abundance
\citep[e.g.,][]{w98}. This will further increase NGC~5018's Mg$_2$ line
strength as it ages so that it may one day fall among other
ellipticals in the Mg$_2$-$\sigma$ relation.

As mentioned previously, significant emission has been observed in the
center of NGC~5018 \citep{ghjn94}.  Since both of
the spectral indices we used depend on the intensity in the center of
a Balmer line, contamination by emission could affect the results
significantly even though the contamination decreases as one moves
toward higher order lines in the Balmer sequence.  Specifically,
emission will weaken H$\delta$ relative to Fe I $\lambda$4045 and to a
lesser degree weaken Ca II H+H$\epsilon$ relative to Ca II K.  Each
index, if contaminated, will therefore have a higher value than if
there were no emission.  From
Figures~\ref{fig:yspm17}--\ref{fig:yspp00}, it can be seen that if
emission contamination were removed, the data points for NGC~5018
would shift toward younger ages.  The determined YSP age of 2.8
Gyr represents therefore an \emph{upper limit on the YSP age}.  The limited
spectral coverage does not permit a more detailed analysis of the
emission contamination, but the fact that H$\beta$ from Figure
\ref{fig:lick} gives the same age indicates that the emission must be
relatively modest.

The power of the method used in this paper is its ability to discriminate
between different, plausible stellar populations.  The verification of the 
presence of a YSP in
NGC~5018 and its age of~2.8 Gyr are quite unambiguous, irrespective of the nature 
of the old, underlying population.  Also, this type of determination is not
unique to NGC~5018.  A similar procedure can be applied to any galaxy with
morphological peculiarities suspected of harboring more than one coeval stellar
population.  With this tool, the effects of a dynamical interaction on the
integrated light of a galaxy can be analyzed more fully.

\acknowledgments

AJL would like to thank Dr.\  Jim Rose for a lot of guidance and many helpful
discussions, Lewis Jones for providing the spectra of NGC 5018 and the referee,
Dr.\  Scott Trager, for many helpful suggestions which improved the paper.  This 
research was partially supported by NSF grant AST-9320723 to the University of 
North Carolina and by NASA through grant GO-06664.01-95A awarded by the Space
Telescope Science Institute which is operated by the Association of
Universities for Research in Astronomy, Inc., for NASA under Contract
NAS5-26555. 

\clearpage

%
%

%

\clearpage

\begin{figure}
\caption{Two age-sensitive indices, a UV-visual color and an
H$\beta$ index, are plotted versus metal-sensitive spectral index
C$_2$4668. The color $\lambda 2750 - V = {\rm log} F_{\lambda 2750} -
{\rm log} F_V$ following Burstein et al. 1988 and BBB, and has units of
dex rather than magnitudes.  The marked dust screen extinction also
refers to a straight logarithm, that is, $A_V = 0.1$ dex, or 0.25 mag. 
The $\lambda 2750$ passband is an average of BBB and Burstein et al.
data from $\lambda 2650$ to $\lambda 2850$ with their extinction
corrections applied. Optical spectral indices on the Lick system come
from Trager et al. (1998). The optical data was taken from a
significantly smaller aperture than the IUE UV data, so the C$_2$4668
index should be weakened somewhat (a few tenths of \AA ) to compensate
for this mismatch in the upper panel. Error bars in bold refer to M32
and M31 data, wider bars to the rest of the sample. Models from Worthey
(1994) appear as a grid labeled by scaled solar metallicity and age in
Gyr. Models are scaled solar and thus do not track element
overenhancements. Such an effect probably exists for the C$_2$4668
index; the larger galaxies have a C$_2$4668 index stronger than their
average abundance would dictate. In the upper panel, an arrow attached
to M31 is an estimate of the effect of subtracting the hot star
component (which is not included in the models) from the spectrum, as
modeled by Worthey et al. (1996). The indicated effect is an
underestimate because Worthey et al. maximized the contribution from
very hot PAGB stars. Similar corrections should be made for NGC 3115 and
NGC 3379, but we lack appropriate spectrophotometric data. NGC 3115 has
a UV upturn similar to M31, but NGC 3379 has considerably more UV flux,
so its correction is likely to be quite large. The other galaxies have
very weak UV emission so that their corrections are negligible. Large
corrections for emission fill-in in the H$\beta$ index are unlikely for
the galaxies plotted. If corrections are applied, galaxies move
vertically to younger ages and somewhat higher abundance.}
\label{fig:lick}
\end{figure}
 
\begin{figure}
\caption{One of the long-slit spectra of NGC~5018 used to calculate the mean 
spectral indices.
Each spectra was smoothed with a gaussian of 0.5  pixels. The absorption features
used in the age-dating technique are identified.} 
\label{fig:spec}
\end{figure}
 
\begin{figure}
\caption{The CaII index is plotted vs the H$\delta$/$\lambda$4045 index for an
instantaneous burst of star formation with [Fe/H] = -0.7 (red solid and 
short-dashed line) and [Fe/H] = 0.0 (blue long-dashed line).  The colored 
lines represent the evolution of the index values as the population ages. Various ages
(in Gyr) have been marked by solid squares.  For clarity, the path from 
0.004 Gyr to 0.5 Gyr on the [Fe/H] = -0.7 curve has been marked with a
short-dashed line while the evolution subsequent to 0.5 Gyr has been marked
with a solid line.  An old, metal-poor model (green circle) is plotted to show the
effects of horizontal branch morphology on the indices (see text). The error bars 
for the labeled globular
cluster points (open triangles) are smaller than the plotting symbol.} \label{fig:burst}
\end{figure}

\begin{figure}
\caption{Ca II plotted vs H$\delta$/$\lambda$4045 for a composite population
consisting of a metal-poor young population ([Fe/H] = -1.7; evolution
marked by the solid black curve, with specific ages labeled by the solid squares) 
and one of two OSP (age = 15.1 Gyr) populations ([Fe/H] = -1.7, marked by the 
solid red circle 
and [Fe/H] = 0.0, marked by the solid blue circle).
Dotted lines between the old population point and the 13.2 Gyr YSP point denote
interpolations between these two populations with each cross along the
OSP [Fe/H] = 0.0 curve marked with the fractional contribution of the YSP
to the total integrated light at 4000 \AA. The crosses along the OSP [Fe/H] = -1.7
curve are similar but have been left off for clarity.   The indices for NGC~5018, 
M32, and 47 Tuc are plotted as open triangles.} \label{fig:yspm17}
\end{figure}

\begin{figure}
\caption{Same as Fig. \ref{fig:yspm17} but with the YSP [Fe/H] = -0.7.} 
 \label{fig:yspm07}
\end{figure}
\begin{figure}
\caption{Same as Fig. \ref{fig:yspm17} but with the YSP [Fe/H] = 0.0. Also,
two other old populations have been added: [Fe/H] = -0.7 (green solid circle)
and [Fe/H] = +0.4 (magenta solid circle).  For clarity, the interpolations between
the OSPs and the YSP have been omitted and are shown in Fig. \ref{fig:yspp00} on
an expanded scale.} \label{fig:yspp00full}
\end{figure}

\begin{figure}
\caption{Identical to Fig. \ref{fig:yspp00full} but with an expanded scale on the
Ca II axis. Also, the interpolations between the OSP points and the YSP points
nearest NGC~5018 are shown.
The increment crosses along the
[Fe/H] = -0.7 and [Fe/H] = +0.0 curves have been omitted for clarity.}
\label{fig:yspp00}
\end{figure}

%


\begin{thebibliography}{}
\bibitem[Bender, Burstein, \& Faber(1993)]{bbf93} Bender, R., Burstein, D., \& Faber, S. M. 1993, \apj, 411, 153
\bibitem[Bertelli et al.(1994)]{bbcfn94} Bertelli, G., Bressan, A., Chiosi, C., Fagotto, F., \&
     Nasi, E., 1994, \aaps, 106, 275
\bibitem[Bertola, Burstein, \& Buson(1993)]{bbb93} Bertola, F., Burstein, D., \& Buson, L. M., 1993, \apj,
     403, 573 (BBB)
\bibitem[Bica, Alloin, \& Schmidt(1990)]{bas90} Bica, E., Alloin, D., \& Schmidt, A., 1990, \aap, 228, 23
\bibitem[Bruzual \& Charlot(1993)]{bc93} Bruzual A., G., \& Charlot, S., 1993, \apj, 405, 538
\bibitem[Burstein et al.(1988)]{bbbfl88} Burstein, D., Bertola, F., Buson, L. M., Faber, S. M., \& Lauer, T. R.
1988, \apj, 328, 440
\bibitem[Bushouse, Lamb, \& Werner(1988)]{blw88} Bushouse, H. A., Lamb, S. A., \& Werner, M. W., 1988, \apj, 335, 74
\bibitem[Carollo \& Danziger(1994)]{cd94} Carollo, C. M., \& Danziger, I. J., 1994, \mnras, 270, 743
\bibitem[Charlot \& Silk(1994)]{cs94} Charlot, S., \& Silk, J., 1994, \apj, 432, 453
\bibitem[Couch \& Sharples(1987)]{cs87} Couch, W. J., \& Sharples, R. M., 1987, \mnras, 229, 423
\bibitem[Fort et al.(1986)]{fpcmv86} Fort, B. P., Prieur, J.-L., Carter, D., Meatheringham, S. J.,
     \& Vigroux, L., 1986, \apj, 306, 110
\bibitem[Goudfrooij et al.(1994)]{ghjn94} Goudfrooij, P., Hansen, L., J\mbox{\o}rgensen, H. E., \& 
     N\mbox{\o}rgaard-Nielsen,  H. U., 1994, \aaps, 105, 341
\bibitem[Helou(1986)]{h86} Helou, G., 1986, \apj, 311, L33
\bibitem[Hernquist \& Quinn(1988)]{hq88} Hernquist, L., \& Quinn, P. J., 1988, \apj, 331, 682
\bibitem[Hilker \& Kissler-Patig(1996)]{hk96} Hilker, M., \& Kissler-Patig, M., 1996, \aap, 314, 357
\bibitem[Jacoby, Hunter, \& Christian(1984)]{jhc84} Jacoby, G. H., Hunter, D. A., \& Christian, C. A., 1984, \apjs,
     56, 257
\bibitem[Jones(1999)]{j99} Jones, L. A., 1999, Ph.D. thesis Univ. of North Carolina
\bibitem[Jura et al.(1987)]{jkkg87} Jura, M., Kim, D.-W., Knapp, G. P., \& Guhathakurta, P., 1987, \apj,
     312, L11
\bibitem[Kim et al.(1988)]{kgvjk88} Kim, D.-W., Guhathakurta, P., van Gorkom, J. H., Jura, M., 
     \& Knapp, G. R., 1988, \apj, 330, 684
\bibitem[Kurucz(1995)]{k95} Kurucz, R. L., 1995, private communication
\bibitem[Lee(1989)]{l89} Lee, Y.-W., 1989, Ph.D. thesis Yale University
\bibitem[Leitherer et al.(1996)]{lea96} Leitherer, C., et al., 1996, \pasp, 108, 996
\bibitem[Leonardi(2000)]{l00} Leonardi, A. J., 2000, in preparation
\bibitem[Leonardi \& Rose(1996)]{lr96} Leonardi, A. J., \& Rose, J. A., 1996, \aj, 111, 182 (LR)
\bibitem[Malin \& Carter(1983)]{mc83} Malin, D. F., \& Carter, D., 1983, \apj, 274, 534
\bibitem[Malin \& Hadley(1997)]{mh97} Malin, D. F., \& Hadley, B., 1997, Pub. Astro. Soc. of Aus., 14, 52
\bibitem[Mihos, Bothun, \& Richstone(1993)]{mbr93} Mihos, J. C., Bothun, G. D., \& Richstone, D. O., 1993, \apj,
     418, 82
\bibitem[Mihos \& Hernquist(1994a)]{mh94a} Mihos, J. C., \& Hernquist, L., 1994a, \apj, 425, L13
\bibitem[Mihos \& Hernquist(1994b)]{mh94b} Mihos, J. C., \& Hernquist, L., 1994b, \apj, 431, L9
\bibitem[O'Connell(1988)]{o88} O'Connell, R. W., 1988, in Towards Understanding Galaxies at High
     Redshift, ed. R. G. Kron \& A. Renzini (Dordrecht: Kluwer), 177
\bibitem[Quinn(1984)]{q84} Quinn, P. J., 1984, \apj, 279, 596
\bibitem[Rose(1984)]{r84} Rose, J. A., 1984, \aj, 89, 1238
\bibitem[Rose(1985)]{r85} Rose, J. A., 1985, \aj, 90, 1927
\bibitem[Schweizer(1983)]{s83} Schweizer, F., 1983, in Proc. IAU Symp. 100, Internal Kinematics
     and Dynamics of Galaxies, ed. E. Athanassoula (Dordrecht: Reidel), 319
\bibitem[Schweizer et al.(1990)]{ssfbdg90} Schweizer, F., Seitzer, P., Faber, S. M., Burstein, D., 
     Dalle Ore, C. M., \& Gonzalez, J. J., 1990, \apj, 364, L33
\bibitem[Thronson \& Bally(1987)]{tb87} Thronson, H. A., \& Bally, J., 1987, \apj, 319, L63
\bibitem[Toomre \& Toomre(1972)]{tt72} Toomre, A., \& Toomre, J., 1972, \apj, 178, 623
\bibitem[Trager et al.(1998)]{twfbg98} Trager, S. C., Worthey, G., Faber, S. M., Burstein, D.,
     \& Gonz\'alez, J. J. 1998, \apjs, 116, 1
\bibitem[Weil \& Hernquist(1993)]{wh93} Weil, M. L., \& Hernquist, L., 1993, \apj, 405, 142
\bibitem[Worthey(1994)]{w94} Worthey, G., 1994, \apjs, 95, 107
\bibitem[Worthey(1998)]{w98} Worthey, G., 1998, \pasp, 110, 888
\bibitem[Worthey, Dorman, \& Jones(1996)]{wdj96} Worthey, G., Dorman, B., \& Jones, L. A. 1996, \aj, 112, 948

\end{thebibliography}
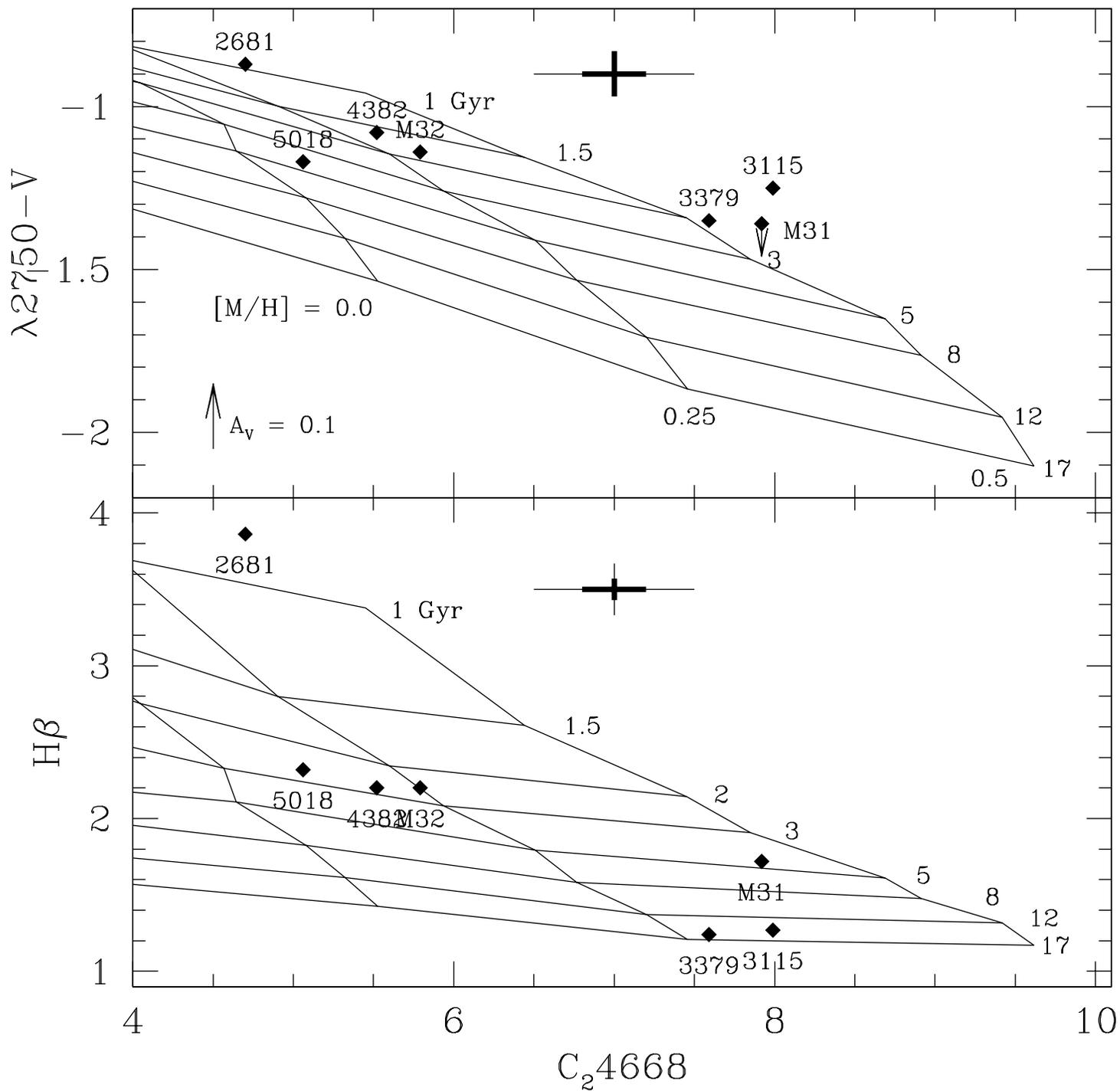
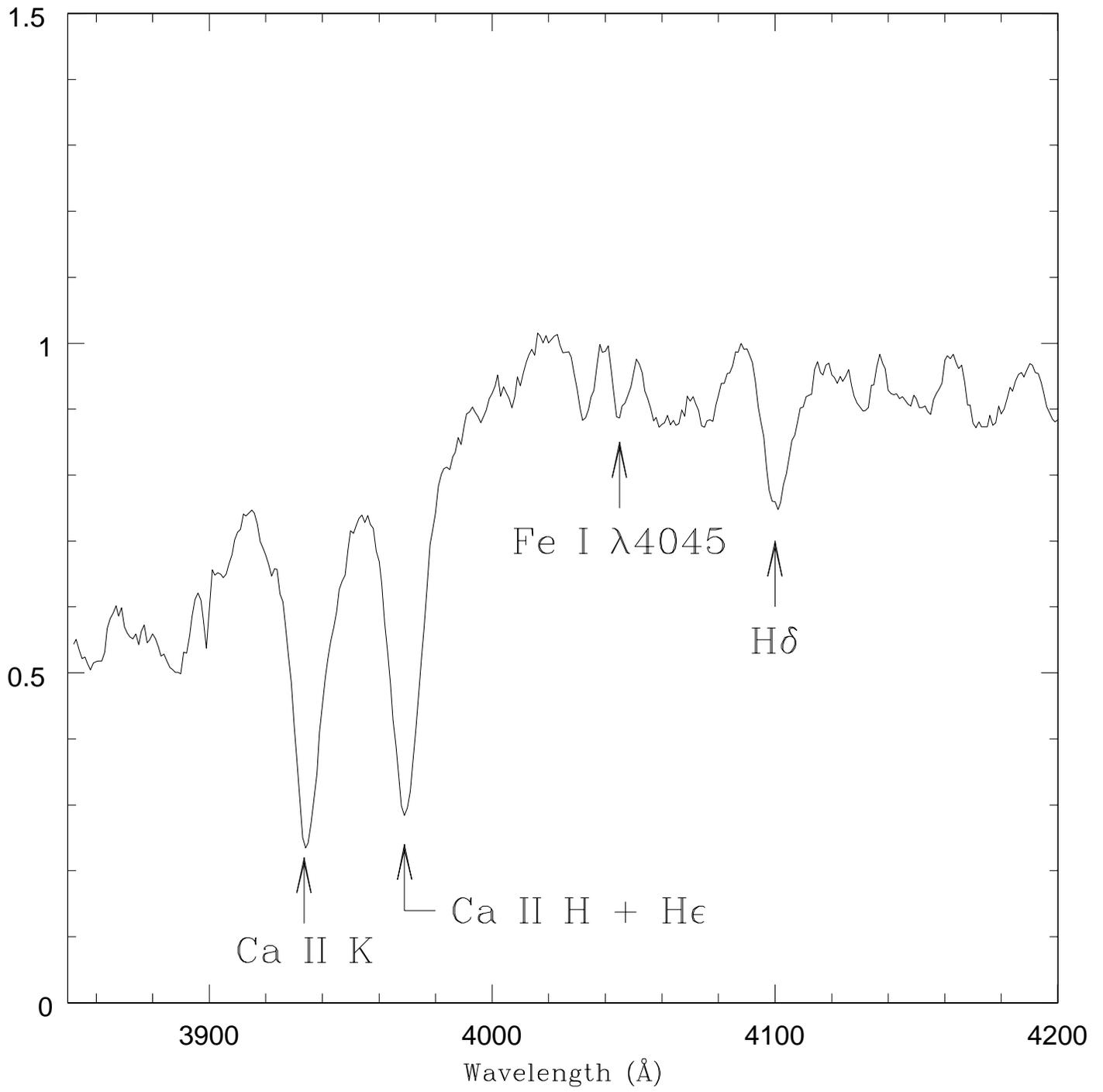
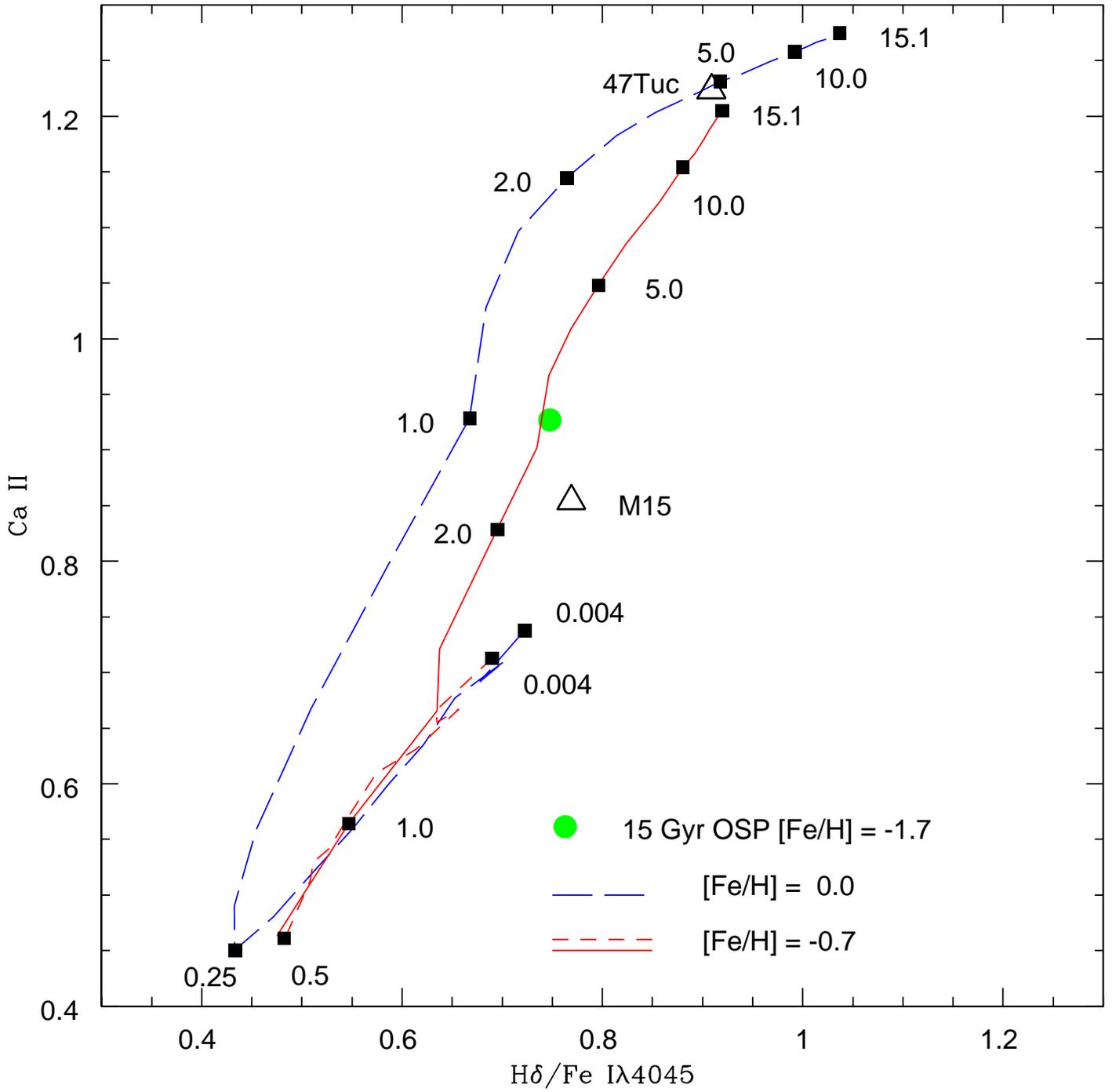
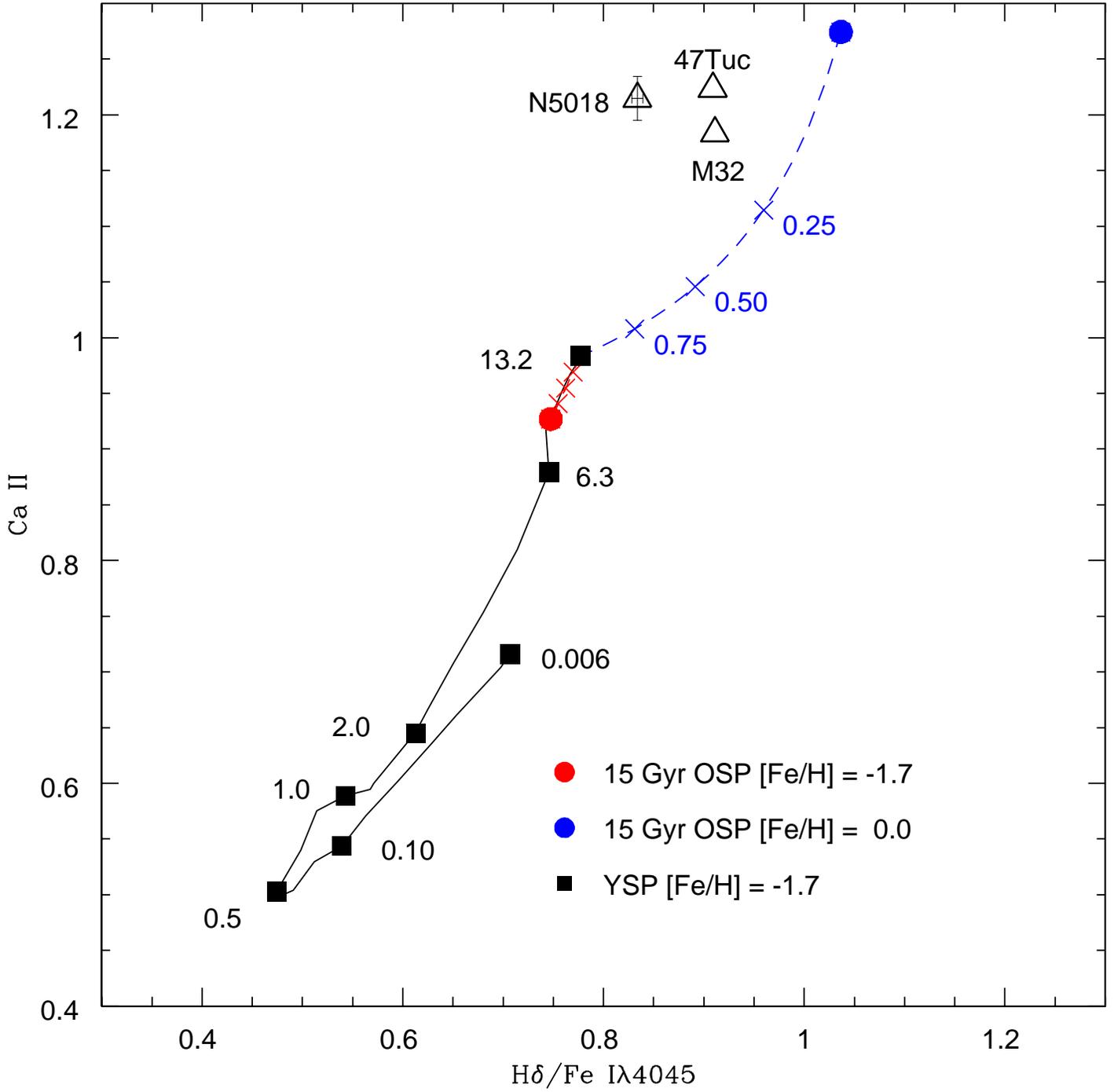
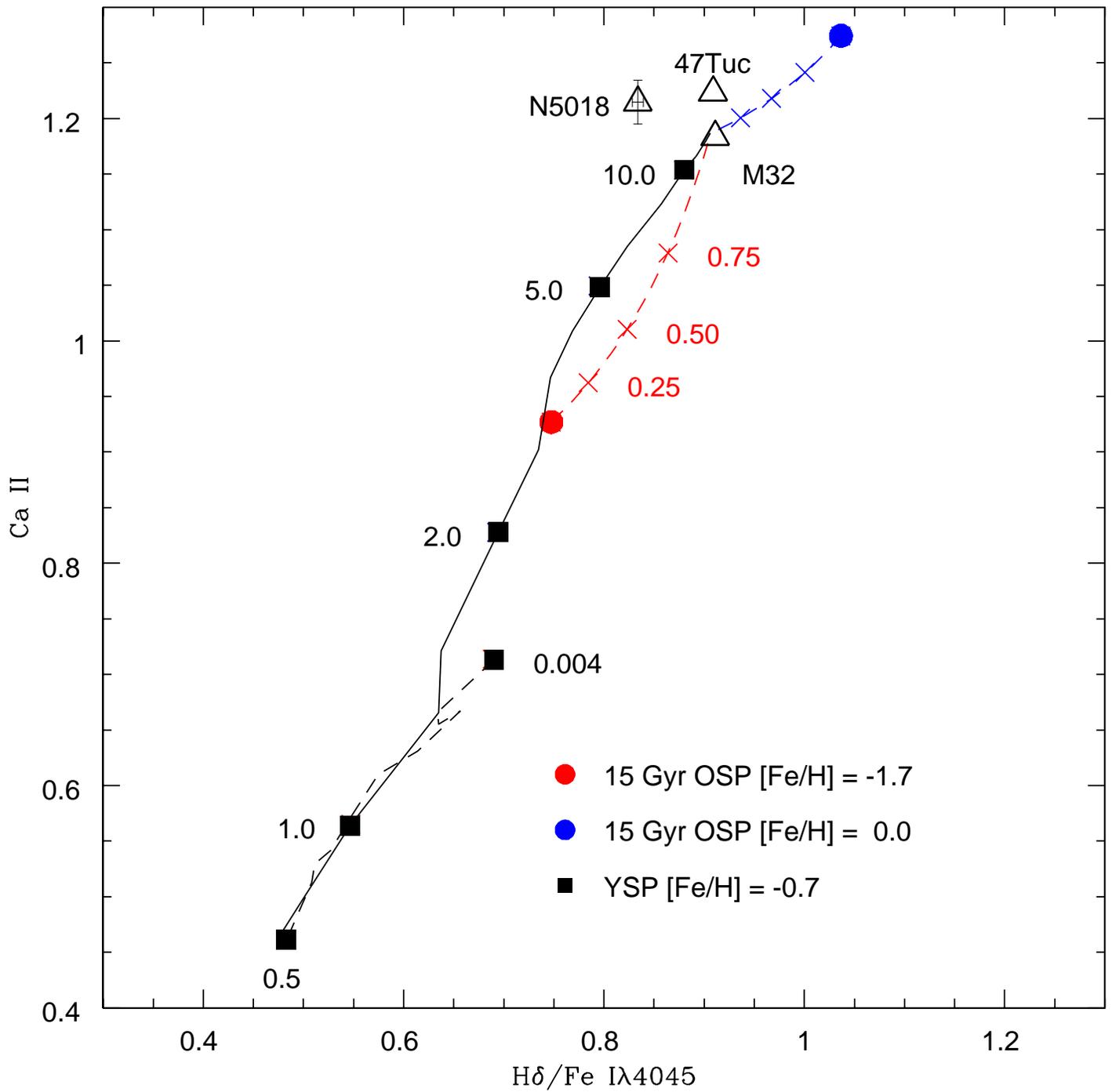
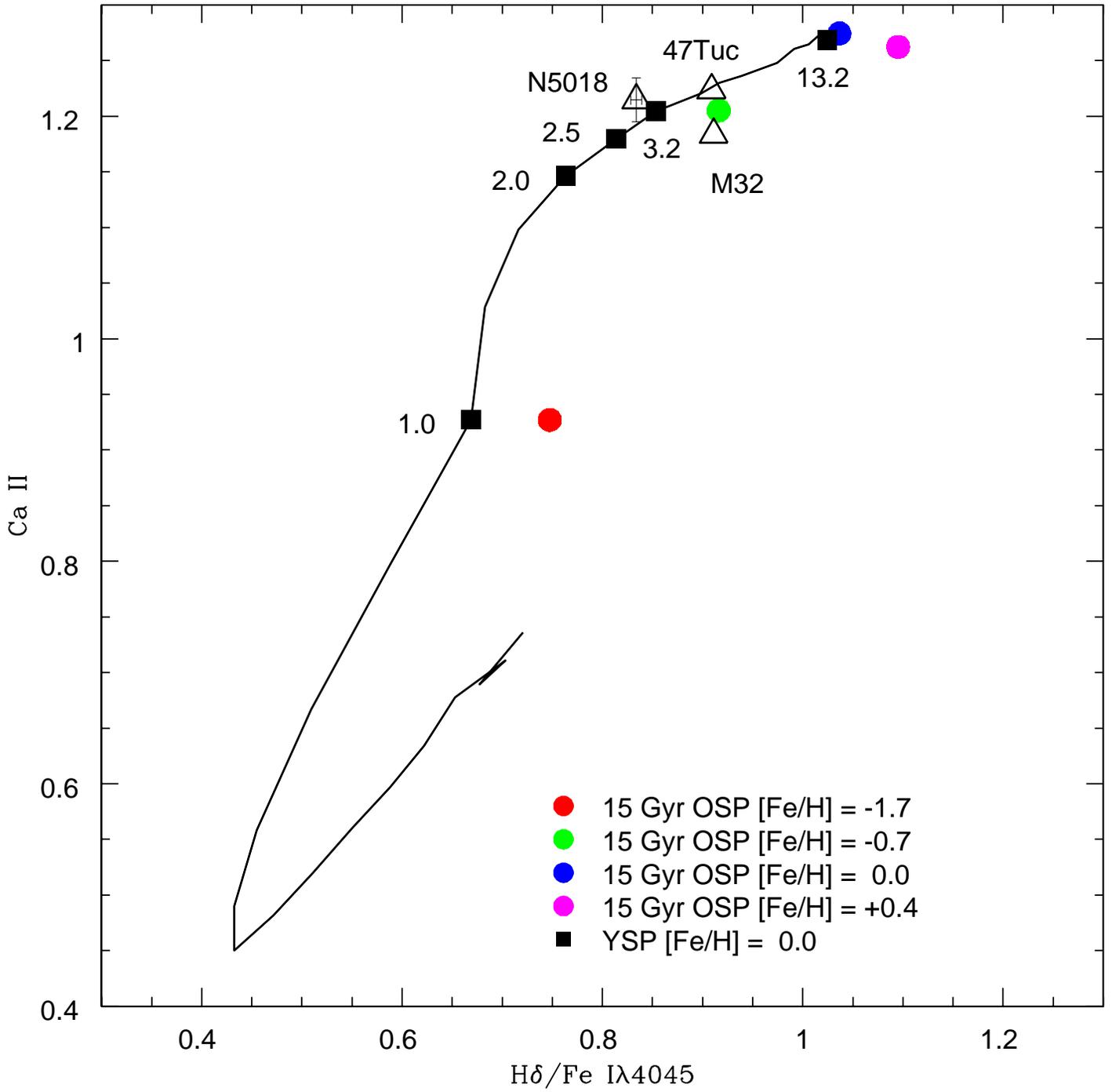
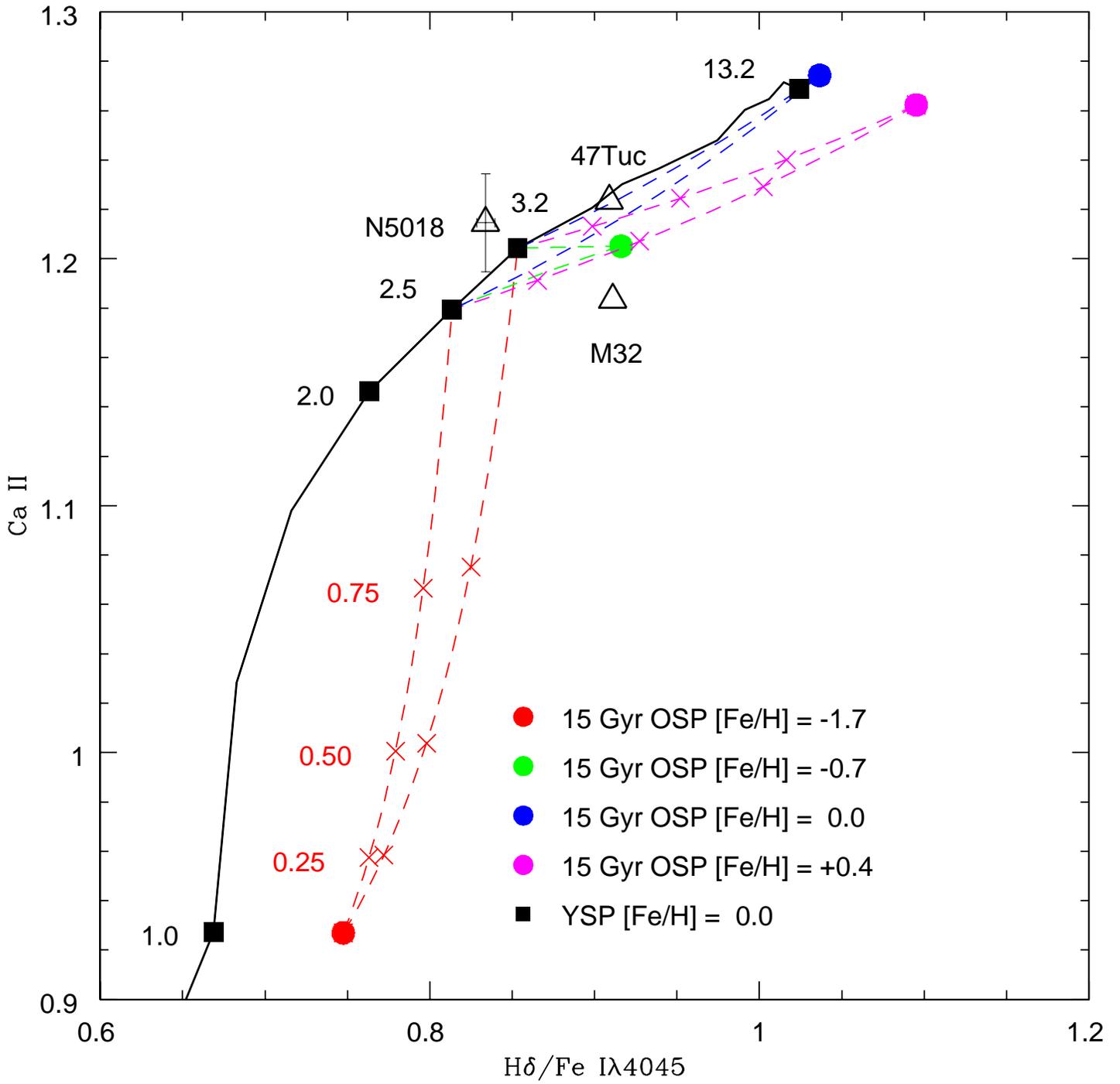
\end{document}